\documentclass[tightenlines,superscriptaddress,eqsecnum,floats,nofootinbib,showpacs]
{revtex4-1}

\usepackage{amssymb}
\usepackage{stmaryrd}
\usepackage{amsmath}
\usepackage{amsfonts}
\usepackage{mathrsfs}
\usepackage{CJK}
\usepackage{amsmath,amssymb,amsfonts}
\usepackage{graphicx}

\def\be{\begin{equation}}
\def\ee{\end{equation}}
\def\ba{\begin{eqnarray}}
\def\ea{\end{eqnarray}}
\def\nn{\nonumber}

\newcommand{\mubar}{{\bar \mu}} 
\newcommand{\abs}[1]{{\left|{#1}\right|}} 
\newcommand{\Abs}[1]{{\Big|{#1}\Big|}} 
\newcommand{\ket}[1]{\vert{#1}\rangle} 
\newcommand{\bra}[1]{\langle{#1}\vert} 
\newcommand{\kt}{{\tilde{K}}} 
\newcommand{\R}{\mathcal {R}} 
\newcommand{\ct}{\tilde{c}}

\newcommand{\sgn}{\mathrm{sgn}} 
\newcommand{\grav}{\mathrm{gr}} 
\newcommand{\sca}{\mathrm{sc}} 
\newcommand{\kin}{\mathrm{kin}} 
\newcommand{\hil}{\mathcal{H}} 

\begin{document}


\title{Loop quantum Brans-Dicke cosmology}

\author{Xiangdong Zhang\footnote{zhangxiangdong@mail.bnu.edu.cn}}
\affiliation{Department of Physics, South China University of
Technology, GuangZhou 510641, China} \affiliation{Department of
Physics, Beijing Normal University, Beijing 100875, China}

\author{Michal Artymowski}
\affiliation{Department of Physics, Beijing Normal University,
Beijing 100875, China}

\author{Yongge Ma\footnote{Corresponding author: mayg@bnu.edu.cn}}
\affiliation{Department of Physics, Beijing Normal University,
Beijing 100875, China}

\begin{abstract}
The spatially flat and isotropic cosmological model of Brans-Dicke theory with coupling parameter
$\omega\neq-\frac{3}{2}$ is quantized by the approach of loop quantum cosmology. An interesting feature of this model is
that, although the Brans-Dicke scalar field is non-minimally coupled with curvature, it
can still play the role of an emergent time variable. In the quantum theory, the classical differential equation
which represents cosmological evolution is replaced by a quantum
difference equation. The effective Hamiltonian and modified dynamical equations of loop quantum Brans-Dicke cosmology are
also obtained, which lay a foundation for the phenomenological investigation to possible quantum gravity effects in cosmology. The effective equations indicate that the classical big bang
singularity is again replaced by a quantum bounce in loop quantum Brans-Dicke
cosmology.

\end{abstract}


\maketitle

\section{Introduction}\label{sec:introduction}

As a background independent
approach to quantize general relativity (GR), loop quantum gravity (LQG) has been widely
investigated in the past 25 years\cite{Ro04,Th07,As04,Ma07}.
Recently, this non-perturbatively loop quantization procedure has
been successfully generalized to the metric $f(\R)$ theories\cite{Zh11,Zh11b},
Brsns-Dicke theory \cite{ZM12a} and scalar-tensor
theories\cite{ZM11c}. In fact, the scheme of these loop quantum modified gravity theories can be extended to
more general metric theories of gravity with well-defined geometrical dynamics \cite{Ma12a}. However, to go round the extreme complexity
of a full theory of quantum gravity, one approach usually
taken is to apply the formal quantization prescriptions to
symmetry-reduced models. These relatively simple toy
models could be employed to test the ideas and constructions of the
full theory and to draw some physical predictions. The so-called loop quantum cosmology (LQC) is such
a symmetry-reduced model from LQG. We refer to \cite{LQC5,Boj,APS3,AS11}
for reviews on LQC. Similarly, to further test the constructions and explore the physical contents of
loop quantum scalar-tensor theories, it is desirable to study their symmetry-reduced models, such as
cosmological models. Among all scalar-tensor theories of gravity, the most simple one is the so-called Brans-Dicke theory
which was introduced Brans and Dicke in 1961 to modify GR in
accordance with Mach's principle \cite{BD}.

The cosmological models of classical Brans-Dicke theory were first studied in
\cite{Greenstein1,Greenstein2}. Then many aspects of Brans-Dicke cosmology have been
widely investigated in the past decades\cite{BD2}. The scalar field non-minimally coupled with curvature in Brans-Dicke theory is even expected to account for the dark energy problem \cite{Banerjee,Sen,Qiang,DB,FT,BD1}, which has become a topical issue in cosmology \cite{01}. It should be noted that the solar system experiments
constrain the coupling constant $\omega$ of the original 4-dimensional Brans-Dicke theory to be a very large number \cite{will,will1}.
For simplicity consideration and consistency with the solar system experiments, we will only consider the original
Brans-Dicke theory with coupling constant $\omega\neq-\frac{3}{2}$.

This paper is organized as follows.
The canonical structure and connection dynamics in the spatially flat FRW model of classical Brans-Dicke theory is first given in section \ref{sec:2}. Then
we construct the loop quantum Brans-Dicke cosmology in section
\ref{section3}, where the dynamical difference equation representing cosmological evolution in the quantum theory is derived. In section \ref{section4},
by simplifying our quantum Hamiltonian constraint, the path
integral method is employed to obtain an effective Hamiltonian constraint. In the light
of this effective Hamiltonian, the effective dynamical equations of loop quantum Brans-Dicke
cosmology is derived in section \ref{section6}, which implies a quantum bounce near to the classical big bang singularity. Conclusions and
outlooks are given in the last section.

\section{canonical structure of Brans-Dicke cosmology}\label{sec:2}

The original gravitational action of 4-dimensional Brans-Dicke theory reads \cite{BD}
\ba
S(g)=\frac{1}{16\pi G}\int_\Sigma
d^4x\sqrt{-g}[\phi\R-\frac{\omega}{\phi}(\partial_\mu\phi)\partial^\mu\phi]\label{action}
\ea
where $\phi$ is a scalar field, $\R$ denotes the scalar curvature of
spacetime metric $g_{\mu\nu}$, and $\omega$ is the coupling constant. Now we consider the spatially flat,
homogeneous and isotropic model.
According to the cosmological principle, one can write the line element of the
spacetime metric of our universe as the following standard form, which is the so-called
Friedman-Robertson-Walker (FRW) metric
\ba
ds^2=-dt^2+a^2(t)\left(dr^2+r^2(d\theta^2+\sin^2\theta d\phi^2)\right) \nn\ea
where $a$ is the scale factor. At classical level, if one assumes that the matter constituent of the universe be some perfect fluid, the evolution
equations of Brans-Dicke cosmology would read \cite{Greenstein1}
\ba \left(\frac{\dot{a}}{a}
+\frac{\dot{\phi}}{2\phi}\right)^2&=&\frac{2\omega+3}{12}\left(\frac{\dot{\phi}}{\phi}\right)^2+\frac{8\pi
G\rho}{3\phi},\label{fried}\\
\frac{\ddot{a}}{a}+2\left(\frac{\dot{a}}{a}\right)^2+\frac{\dot{a}\dot{\phi}}{a\phi}&=&\frac{8\pi
G}{(3+2\omega)\phi}\left(-\omega P+(\omega+1)\rho\right),\ea
and the
equation of motion for the scalar field is
\ba
-\frac{1}{a^3}\frac{d}{dt}(\dot{\phi}a^3)&=&\frac{8\pi
G}{3+2\omega}(-\rho+3P),\nn\ea
where a dot over a letter denotes the derivative with respect to the cosmological time $t$, $\rho$ and $P$ are respectively the energy density and pressure of the fluid.
In the case that the matter part is a massless scalar field,
because $P=\rho$, the above equation will reduce to
\ba
-\frac{1}{a^3}\frac{d}{dt}(\dot{\phi}a^3)&=&\frac{16\pi
G}{3+2\omega}\rho. \label{seom}\ea

Recall that loop quantum scalar-tensor theories are based on their connection dynamical formalism \cite{ZM11c}, where the phase
space consists of canonical pairs of geometrical conjugate variables, $SU(2)$ connection
$A_a^i$ and densitized triad $E^b_j$, and scalar conjugate variables $(\phi,\pi)$. The Poisson brackets between the
canonical variables read
\ba
\{A^j_a(x),E_k^b(y)\}&=&\kappa\gamma\delta^b_a\delta^j_k\delta(x,y),\nn\\
\{\phi(x),\pi(y)\}&=&\delta(x,y), \nn\ea
where $\kappa=8\pi G$. To mimic the full theory, we can do the following symmetric reduction of the connection formalism as in standard LQC.
We first introduce an ``elemental cell" $\mathcal {V}$ on the homogeneous spatial manifold $\mathbb{R}^3$ and restrict all
integrals to this elemental cell. Then we choose a fiducial Euclidean metric ${}^oq_{ab}$ on $\mathbb{R}^3$ as well as the orthonormal triad and co-triad
$({}^oe^a_i ; {}^o\omega^i_a)$, such that
${}^oq_{ab}={}^o\omega^i_a{}^o\omega^i_b$.
For simplicity, we let the
 elemental cell $\mathcal {V}$ be cubic as measured by
${}^oq_{ab}$ and denote its volume by $V_o$. For spatially flat FRW model we have $A_a^i=\gamma
\kt_a^i$, where $\gamma$ is a nonzero real number and $\kt_a^i$ is defined in \cite{ZM11c}. Via fixing the degrees of freedom of local gauge and
diffeomorphism transformations, we finally yield the reduced connection and densitized
triad as \cite{LQC5}
\ba A_a^i=\ct
V_0^{-\frac13}\ {}^o\omega^i_a,\quad\quad\quad
E^b_j=pV_0^{-\frac23}\sqrt{\det({}^0q)}\ {}^oe^b_j, \nn\ea
where $\ct,p$
are only functions of $t$. Hence the phase space of
cosmological model consists of conjugate pairs $(\ct,p)$ and
$(\phi,\pi)$. The Poisson brackets between them read
\ba
\{\ct,p\}&=&\frac{\kappa}{3}\gamma,\nn\\
\{\phi,\pi\}&=&1. \label{poissonb}\ea
Note that the new variables are related to the old ones by $\abs{p}=a^2V_0^{\frac 23}$ and
$\ct=(\phi\dot{a}+\frac{a}{2}\dot{\phi})\gamma V_0^{\frac 13}$.

The Gaussian and diffeomorphism constraints in the full theory have been solved by the symmetric reduction. Hence, in the cosmological model we only need to treat the
remaining Hamiltonian constraint. Its expression in the full theory reads \cite{ZM12a}
\ba
H&=&\frac{\phi}{2\kappa}\left[F^j_{ab}-(\gamma^2+\frac{1}{\phi^2})\varepsilon_{jmn}\tilde{K}^m_a\tilde{K}^n_b\right]\frac{\varepsilon_{jkl}
E^a_kE^b_l}{\sqrt{q}}\nn\\
&+&\frac{\kappa}{3+2\omega}\left(\frac{(\tilde{K}^i_aE^a_i)^2}{\kappa^2\phi\sqrt{q}}+
2\frac{(\tilde{K}^i_aE^a_i)\pi}{\kappa\sqrt{q}}+\frac{\pi^2\phi}{\sqrt{q}}\right) \nn\\
&+&\frac{\omega}{2\kappa\phi}\sqrt{q}(D_a\phi)
D^a\phi+\frac{1}{\kappa}\sqrt{q}D_aD^a\phi \nn\\
&=&0.\label{BDhamilton} \ea
In the cosmological model which we are considering,
the above Hamiltonian constraint reduces to
\ba
H&=&-\frac{3\ct^2\sqrt{\abs{p}}}{\gamma^2\kappa\phi}+\frac{\kappa}{(3+2\omega)\phi
\abs{p}^{\frac32}}(\frac{3\ct p}{\kappa\gamma}+\pi\phi)^2=0.\label{Chamilton}\ea

Recall that in the cosmological model of GR minimally coupled with a massless scalar field, the scalar field can be viewed as an emergent internal time variable. An interesting question arising in our Brans-Dicke cosmology is that whether the scalar field nonminimally coupled with the geometry can still be viewed as emergent time. To answer this
question, we check the evolution equation of the scalar field,
\ba
\dot{\phi}=\{\phi,H\}=\frac{2\kappa}{(3+2\omega)\abs{p}^{\frac32}}(\frac{3\ct
p}{\kappa\gamma}+\pi\phi). \nn\ea
If we could show that
$p_{\phi}=\frac{3\ct p}{\kappa\gamma}+\pi\phi$ is a constant of
motion, the scalar field would be a monotonic
function with respect to the cosmological time. This is indeed the case since
\ba \dot{p_{\phi}}=
\{p_{\phi},H\}
&=&-\frac{1}{2}(-\frac{6\omega\ct^2\sqrt{\abs{p}}}{(3+2\omega)\kappa\gamma^2\phi}+\frac{6\ct\pi\sgn(p)}{(3+2\omega)\gamma\sqrt{\abs{p}}}
+\frac{\kappa\pi^2\phi}{(3+2\omega)\abs{p}^{\frac32}})\nn\\
&=&-\frac{1}{2}H\approx0, \nn\ea
where $\sgn(p)$ is the sign function for
$p$. Therefore we conclude that, although the scalar field is
nonminimally coupled with geometry in Brans-Dicke cosmology, it can still be viewed as an
emergent time variable.

\section{Loop Quantization of Brans-Dicke cosmology} \label{section3}

To quantize the cosmological model, we first need to construct the
quantum kinematics of Brans-Dicke cosmology by mimicking the loop
quantum scalar-tensor theory. This is the so-called polymer-like quantization. The
kinematical Hilbert space for the geometry part can be defined as
$\mathcal{H}_{\kin}^{\grav}:=L^2(R_{Bohr},d\mu_{H})$, where
$R_{Bohr}$ and $d\mu_{H}$ are respectively the Bohr
compactification of the real line and Haar measure on it
\cite{LQC5}. On the other hand, for convenience we choose
Schrodinger representation for the scalar field \cite{AS11}. Thus
the kinematical Hilbert space for the scalar field part is defined as
in usual quantum mechanics,
$\mathcal{H}_{\kin}^{\sca}:=L^2(R,d\mu)$. Hence the whole Hilbert
space of Brans-Dicke cosmology is a direct product, $\hil_\kin^{BD}=\hil^\grav_\kin\otimes
\hil^\sca_\kin$. Now let $\ket{\mu}$ be the eigenstates of
 $\hat{p}$ in the kinematical Hilbert space $\mathcal{H}_{\kin}^{\grav}$ such that
 \ba
\hat{p}\ket{\mu}=\frac{8\pi G\gamma\hbar}{6}\mu\ket{\mu}. \nn\ea
Then those eigenstates satisfy orthonormal condition
\ba
\bra{\mu_i}{\mu_j}\rangle=\delta_{\mu_i,\mu_j}\ , \ea
where $\delta_{\mu_i,\mu_j}$ is the Kronecker delta function rather than the Dirac distribution.
For the convenience of studying quantum dynamics, we define new
variables
\ba v:=2\sqrt{3}sgn(p)\mubar^{-3},\quad b:=\mubar \ct, \nn\ea
where
$\mubar=\sqrt{\frac{\Delta}{|p|}}$ with
$\Delta=4\sqrt{3}\pi\gamma{\ell}_{\textrm{p}}^2$ being a minimum
nonzero eigenvalue of the area operator \cite{Ash-view}. They also
form a pair of conjugate variables as
\ba \{b,v\}=\frac{2}{\hbar}\ .\nn
\ea
It turns out that the eigenstates of
 $\hat{v}$ also contribute an orthonormal basis in $\mathcal{H}_{\kin}^{\grav}$.
We denote
$\ket{\phi,v}$ as the generalized orthonormal basis for the whole Hilbert space
$\hil^{BD}_\kin$.
In $(b,v)$ representation, the Hamiltonian constraint (\ref{Chamilton}) can be written
as
\ba
H=-\frac{\sqrt{3\Delta}b^2\abs{v}}{2\kappa\gamma^2\phi}+\frac{2\sqrt{3}\kappa}{(3+2\omega)\phi(\Delta)^{\frac32}
\abs{v}}(\frac{3\hbar bv}{4}+\pi\phi)^2=0.\nn\ea

Now we come to the quantum dynamics. As in usual LQC, we start with the Hamiltonian constraint of full theory. However, since what we consider here is the homogeneous universe, the last
two terms containing spatial derivative in the Hamiltonian (\ref{BDhamilton}) can be neglected. Hence we write Eq.(\ref{BDhamilton}) as $H=\sum^5_{i=1}H_i$. The quantization of the first two terms in $H$ is similar to that in usual LQC
\cite{APS3}. Thus the sum of the first two terms act on a quantum state
$\Psi(\nu,\phi)\in\hil_\kin^{BD}$ as
\ba
(\hat{H}_1+\hat{H}_2)\Psi(\nu,\phi)=\frac1\phi (\sin b) A(v)(\sin
b)\Psi(\nu,\phi)\nn\ea
where
\ba
A(v)\Psi(\nu,\phi)=\frac{\sqrt{3\Delta}}{4\kappa\gamma^2}\abs{v}\abs{\abs{v+1}-\abs{v-1}}\Psi(\nu,\phi).\nn
\ea
Hence the final result is
\ba
(\hat{H}_1+\hat{H}_2)\Psi(\nu,\phi)=\frac1\phi\left(f_+(v)\Psi(\nu+4,\phi)+f_0(v)\Psi(\nu,\phi)+f_-(v)\Psi(\nu-4,\phi)\right),\nn
\ea
where
\ba
f_+(v)=\frac{\sqrt{3\Delta}}{16\kappa\gamma^2}\Abs{\abs{v+3}-\abs{v+1}}\abs{v+2},\nn\\
f_-(v)=f_+(v-4), \quad  f_0(v)=-f_+(v)-f_-(v).\nn\ea
Now we turn to $H_3,H_4,H_5$ terms.
Note that here we need to quantize the term $\kt_a^iE^a_i$.
Due to the spatial flatness, we have
$\kt_a^iE^a_i=\frac{1}{\gamma}A_a^iE^a_i$. In the cosmological
model, this term can be reduced by
\ba
\frac{1}{\gamma}A_a^iE^a_i\rightsquigarrow \frac{3}{\gamma}\ct
p=\frac{3\kappa\hbar bv}{4}.\nn \ea
Because we use polymer
representation for geometry, there is no quantum operator
corresponding to connection $\ct$ as in standard LQC \cite{AS11}.
Hence we have to replace the connection by holonomy to get a
well-defined operator. This can be achieved by using the classical
identity
\ba \lim_{\mubar\rightarrow
0}\frac{h^{(2\mubar)}_i-h^{(2\mubar)^{-1}}_i}{4\mubar}
=\lim_{\mubar\rightarrow
0}\frac{\sin(\mubar\ct)}{\mubar}\tau_i=\ct\tau_i\ ,\nn\ea
where $\tau_i=-\frac{i}{2}\sigma_i$, $\sigma_i$ is Pauli matrices and the
holonomy is defined by
\ba
h^{(\mubar)}_i:=\cos\frac{\mubar\ct}{2}\textit{1}+2\sin\frac{\mubar\ct}{2}\tau_i\ .
\label{holonomy}\ea here $\textit{1}$ is the $2\times 2$ unit matrix.
Thus, according to Eq.(\ref{holonomy}) we can
replace connection $\ct$ by holonomy
$\frac{\sin(\mubar\ct)}{\mubar}$. Then the symmetry-reduced expression of the sum of $H_3,H_4,H_5$ terms becomes
\ba \frac{2\sqrt{3}\kappa}{(3+2\omega)\phi(\Delta)^{\frac32}
\abs{v}}(\frac{3\hbar
bv}{4}+\pi\phi)^2\rightsquigarrow\frac{2\sqrt{3}\kappa}{(3+2\omega)(\Delta)^{\frac32}}\left((\frac{3\hbar}{4})^2\frac{\sin(b)\abs{v}\sin(b)}
{\phi}+2sgn(p)(\frac{3\hbar}{4})\sin(b)\pi+\frac{\pi\phi\pi}{\abs{v}}\right). \nn\ea
Let $\beta=3+2\omega$. Based on the above discussion, the
action of $\hat{H}_3$ on a quantum state read
\ba
\hat{H}_3\Psi(\phi,v)&=&\frac{2\sqrt{3}\kappa}{\beta\phi(\Delta)^{\frac32}}\left(\frac{3\hbar
}{4}\right)^2
\sin(b)\hat{\abs{v}}\sin(b)\Psi(\phi,v)\nn\\
&=&-\frac{\sqrt{3}\kappa}{2\beta\phi(\Delta)^{\frac32}}\left(\frac{3\hbar
}{4}\right)^2\left[\abs{v+2}\Psi(\phi,v+4)-2\abs{v}\Psi(\phi,v)+\abs{v-2}\Psi(\phi,v-4)\right].\nn\ea
Similarly, for $H_4$ term we have
\ba
\hat{H}_4\Psi(\phi,v)&=&\frac{2\sqrt{3}\kappa}{\beta(\Delta)^{\frac32}}\left(\frac{3\hbar
}{4}\right)2sgn(p)\sin(b)\hat{\pi}\Psi(\phi,v)\nn\\
&=&\frac{2\sqrt{3}\kappa}{\beta(\Delta)^{\frac32}}\left(\frac{3\hbar
}{4}\right) \hbar sgn(p)[\frac{\partial\Psi(\phi,v+2)}{\partial\phi}
-\frac{\partial\Psi(\phi,v-2)}{\partial\phi}]. \nn\ea
Also the action of $\hat{H}_5$ takes the
form
\ba
\hat{H}_5\Psi(\phi,v)&=&\frac{2\sqrt{3}\kappa}{\beta(\Delta)^{\frac32}}
\widehat{\abs{v}^{-1}}\hat{\pi}\hat{\phi}\hat{\pi}\Psi(\phi,v)\nn\\
&=&-\frac{2\sqrt{3}\kappa}{\beta(\Delta)^{\frac32}}(\hbar)^2B(v)\frac{\partial}{\partial\phi}\left(\phi\frac{\partial\Psi(\phi,v)}
{\partial\phi}\right),\nn\ea
where
\ba
B(v)=(\frac32)^3\abs{v}\abs{\abs{v+1}^{1/3}-\abs{v-1}^{1/3}}^3. \ea
Thus, the Hamiltonian constraint (\ref{BDhamilton}) has been successfully quantized in the cosmological model. The Hamiltonian constraint equation of loop quantum Brans-Dicke cosmology reads
\ba
(\sum^5_{i=1}\hat{H}_i)\Psi(\phi,v)=0.\label{hbd}
\ea
With this quantum dynamical equation in hand, in principle one can study the evolutional behavior of the universe around the classical big bang singularity by numerical simulation. However, since the terms related to the Brans-Dicke scalar field are involved in the quantum Hamiltonian, the numerical simulation becomes very complicated, which we would like to leave for future study. In this paper, we will employ an effective Hamiltonian instead of the full quantum Hamiltonian to study the dynamical evolution of the universe in this model.
To study the effective
theory of loop quantum Brans-Dicke cosmology, we also want to know
the effect of matter fields on the dynamical evolution. Hence we now
include an extra massless scalar matter field $\varphi$ into
Brans-Dicke cosmology. Then classically the total Hamiltonian
constraint of the model reads
\ba H
&=&-\frac{\sqrt{3\Delta}b^2\abs{v}}{2\kappa\gamma^2\phi}+\frac{2\sqrt{3}\kappa}{(3+2\omega)\phi(\Delta)^{\frac32}
\abs{v}}(\frac{3\hbar
bv}{4}+\pi\phi)^2+\frac{\sqrt{3}p_\varphi^2}{\abs{v}(\Delta)^{\frac32}}=0\label{hcm}\ea
where $p_\varphi$ is the momentum conjugate to $\varphi$. In the quantum theory, the whole
Hilbert space now is a direct product of three parts,
$\hil_\kin^{total}=\hil^\grav_\kin\otimes
\hil^\sca_\kin\otimes\hil^{matter}_\kin$. Here the kinematical Hilbert
space for the matter part is also defined as in usual quantum
mechanics as $\mathcal{H}_{\kin}^{matter}:=L^2(R,d\mu)$.
The action of the Hamiltonian of matter field on a quantum state
$\Psi(v,\phi,\varphi)\in\hil_\kin^{total}$ reads
\ba
\frac{\sqrt{3}\hat{p}_\varphi^2}{(\Delta)^{\frac32}}\widehat{\abs{v}^{-1}}\Psi(v,\phi,\varphi)=-\frac{\sqrt{3}}{(\Delta)^{\frac32}}\hbar^2B(v)
\frac{\partial^2\Psi(v,\phi,\varphi)}{\partial\varphi^2}.\label{hm} \ea

\section{Effective Hamiltonian of Brans-Dicke cosmology}\label{section4}

To test the robustness of key features of loop quantum cosmology, a
simplified soluble model of LQC was proposed in Ref.\cite{ACS}. In
the simplified model, one first adapted the classical theory to the
scalar matter field time by the following "harmonic gauge" form of
spacetime metric:
\ba
ds^2=-a^6(\tau)d\tau^2+a^2(\tau)\left(dr^2+r^2(d\theta^2+\sin^2\theta
d\phi^2)\right). \nn\ea
Then the Hamiltonian constraint (\ref{hcm})
becomes
\ba H_s
&=&-\frac{\Delta^2b^2v^2}{2\kappa\gamma^2\phi}+\frac{2\kappa}{(3+2\omega)\phi}(\frac{3\hbar
bv}{4}+\pi\phi)^2+p_\varphi^2=0.\label{shcm}\ea
In the corresponding
quantum theory, we denote quantum state
$\Psi(v)\equiv\Psi(v,\phi,\varphi)$ for short. Then the simplified
Hamiltonian constraint equation reads
\ba
\frac{\partial^2\Psi(v)}{\partial\varphi^2}=-\hat{\Theta}\Psi(v),\label{hs}\ea
where
\ba
\hat{\Theta}\Psi(v)&=&\frac{\Delta^2}{8\kappa\gamma^2\hbar^2\phi}v\left[(v+2)\Psi(v+4)-2v\Psi(v)+(v-2)\Psi(v-4)\right]\nn\\
&-&\frac{\kappa}{2\beta\hbar^2\phi}\left(\frac{3\hbar}{4}\right)^2
v\left[(v+2)\Psi(v+4)-2v\Psi(v)+(v-2)\Psi(v-4)\right]\nn\\
&+&\frac{2\kappa}{i\beta\hbar^2}\left(\frac{3\hbar}{4}\right) \hat{\pi}
v\left[\Psi(v+2)-\Psi(v-2)\right]
+\frac{2\kappa}{\beta\hbar^2}\hat{\pi}\phi\hat{\pi}\Psi(v)\nn\\
&\equiv&(\sum^4_{i=1}\hat{\Theta}_i)\Psi(v).\label{hamilton}\ea
Thus we get a Klein-Gordon type equation for the quantum dynamics of Brans-Dicke cosmology coupled with a massless scalar field.
Note that the constraint equation (\ref{hs}) could also be reduced from the Hamiltonian (\ref{hbd}) and (\ref{hm}) by the replacements
\cite{ACS}:
\ba B(v)\longmapsto \frac{1}{\abs{v}},\nn
\ea
and
\ba A(v)=\frac{\sqrt{3\Delta}}{2\kappa\gamma^2}\abs{v}. \nn\ea
The first replacement amounts to
assuming $\mathcal {O}(\frac{1}{\abs{v}})\ll 1$, which then implies the validity of the second
replacement.

The effective description of LQC is a delicate and topical issue
since it may relate the quantum gravity effects to low-energy
physics. The effective Hamiltonian of LQC are being studied from both
canonical perspective\cite{Taveras,DMY,YDM,Boj11} and path integral
perspective\cite{ACH102,QHM,QDM,QM1,QM2}.
With the help of
the Hamiltonian constraint equation (\ref{hs}), we now derive an effective
Hamiltonian within the timeless path integral formalism.
In timeless path integral formalism of our model, the transition amplitude equals to the physical inner product \cite{ACH102,QHM}, i.e.,
\begin{align}
A_{tls}(v_f, \phi_f, \varphi_f;~v_i,\phi_i, \varphi_i)=\langle v_f,
\phi_f,\varphi_f|v_i,\phi_i,\varphi_i\rangle_{phy}=\lim\limits_{\alpha_o\rightarrow\infty}
\int_{-\alpha_o}^{\alpha_o}d\alpha\langle
v_f,\phi_f,\varphi_f|e^{i\alpha\hat{C}}|v_i,\phi_i,\varphi_i\rangle, \label{amplitude}
\end{align}
where $\hat{C}\equiv\hat{\Theta}+\hat{p}_{\varphi}^2/\hbar^2$. As shown in Refs.\cite{QHM,QDM}, by multiple group averaging and complete basis inserting, we will need to calculate
\begin{align}
\langle v_f, \phi_f, \varphi_f|e^{i\sum\limits_{n=1}^N{\epsilon\alpha_n}\hat{C}}|v_i,\phi_i,\varphi_i \rangle=\sum\limits_{v_{N-1},...v_1}\int d\phi_{N-1}...d\phi_1\int d\varphi_{N-1}...d\varphi_1\prod\limits_{n=1}^N\langle \varphi_n|\langle \phi_n|\langle v_n|e^{i\epsilon\alpha_n\hat{C}}|v_{n-1}\rangle|\phi_{n-1}\rangle|\varphi_{n-1}\rangle.
\label{insert basis}
\end{align}
Since the action of the constraint operator $\widehat{C}$ has been separated into gravitational part and matter part, we could calculate the exponential on each kinematical space separately. Then,
for the matter part one gets
\begin{align}
\langle{\varphi_n}|e^{i\epsilon\alpha_n\frac{\widehat{p}^2_\varphi}{\hbar^2}}|\varphi_{n-1}\rangle
=&\int dp_{\varphi_n}\langle{\varphi_n}|p_{\varphi_n}\rangle\langle p_{\varphi_n}|e^{i\epsilon\alpha_n\frac{\widehat{p}^2_\varphi}{\hbar^2}}|\varphi_{n-1}\rangle\nonumber\\
=&\frac{1}{2\pi\hbar}\int dp_{\varphi_n}e^{i\epsilon(\frac{p_{\varphi_n}}{\hbar}\frac{\varphi_n-\varphi_{n-1}}{\epsilon}
+\alpha_n\frac{{p}^2_{\varphi_n}}{\hbar^2})}.
\label{material amplitude}
\end{align}
For the gravity part, we first use the following identity
\ba \int
d\phi_{n}\bra{\phi_n}\bra{v_n}e^{-i\epsilon\alpha_n\hat{\Theta}}\ket{v_{n-1}}\ket{\phi_{n-1}}=
\delta_{v_n,v_{n-1}}-i\epsilon\alpha_n\sum_i\int
d\phi_{n}\bra{\phi_n}\bra{v_n}\hat{\Theta}_i\ket{v_{n-1}}\ket{\phi_{n-1}}+\mathcal
{O}(\epsilon^2). \label{GHamilton}\ea
Then, the matrix elements of $\hat{\Theta}_i$
can be calculated separately by using Eq.(\ref{hamilton}). We have
\ba
&&\int d\phi_n\bra{\phi_n}\bra{v_n}\hat{\Theta}_1\ket{v_{n-1}}\ket{\phi_{n-1}}\nn\\
&=&
\frac{\Delta^2}{8\kappa\gamma^2\hbar^2\phi_{n-1}}v_{n-1}\frac{v_n+v_{n-1}}{2}
(\delta_{v_n,v_{n-1}+4}-2\delta_{v_n,v_{n-1}}+\delta_{v_n,v_{n-1}-4})\nn\\
&=&\frac{1}{2\pi\hbar}\int d\phi_{n}
d\pi_{n}e^{i\epsilon(\frac{\pi_{n}}{\hbar}\frac{\phi_n-\phi_{n-1}}{\epsilon})}\frac{\Delta^2}{8\kappa\gamma^2\hbar^2\phi_n}v_{n-1}
\frac{v_n+v_{n-1}}{2}
(\delta_{v_n,v_{n-1}+4}-2\delta_{v_n,v_{n-1}}+\delta_{v_n,v_{n-1}-4}),\nn
\ea
where $\pi_n$ is the momentum conjugate to $\phi_n$. Similarly, we can get
\ba
&&\int d\phi_n\bra{\phi_n}\bra{v_n}\hat{\Theta}_2\ket{v_{n-1}}\ket{\phi_{n-1}}\nn\\
&=&-\frac{\kappa}{2\beta\hbar^2}\left(\frac{3\hbar}{4}\right)^2\frac{1}{2\pi\hbar}\int
d\phi_{n}d\pi_{n}e^{i\epsilon(\frac{\pi_{n}}{\hbar}\frac{\phi_n-\phi_{n-1}}{\epsilon})}\frac{1}{\phi_{n-1}}
v_{n-1}\frac{v_{n}+v_{n-1}}{2}\Big[\delta_{v_n,v_{n-1}+4}-2\delta_{v_n,v_{n-1}}+\delta_{v_n,v_{n-1}-4}\Big],\nn
\ea
and
\ba &&\int
d\phi_n\bra{\phi_n}\bra{v_n}\hat{\Theta}_3\ket{v_{n-1}}\ket{\phi_{n-1}}\nn\\
&&=\frac{2\kappa}{i\beta\hbar^2}
\left(\frac{3\hbar}{4}\right) \frac{1}{2\pi\hbar}\int
d\phi_{n}d\pi_{n}e^{i\epsilon(\frac{\pi_{n}}{\hbar}\frac{\phi_n-\phi_{n-1}}{\epsilon})}\pi_n v_{n-1}
(\delta_{v_n,v_{n-1}+2}-\delta_{v_n,v_{n-1}-2}). \nn\ea
At last we have
\ba
\int d\phi_n\bra{\phi_n}\hat{\Theta}_4\ket{\phi_{n-1}}
&=&\int d\phi_n\bra{\phi_n}\frac{\kappa}{\beta\hbar^2}\left(\hat{\phi}\hat{\pi}^2+\hat{\pi}^2\hat{\phi}\right)\ket{\phi_{n-1}}\nn\\
&=&\frac{\kappa}{\beta\hbar^2}\int d\phi_n\left(\int d\pi_n\bra{\phi_n}\ket{\pi_n}\bra{\pi_n}\hat{\pi}^2\hat{\phi}\ket{\phi_{n-1}}+\int d\pi_n\bra{\phi_n}\hat{\phi}
\hat{\pi}^2\ket{\pi_n}\bra{\pi_n}\ket{\phi_{n-1}}\right)\nn\\
&=&\frac{\kappa}{\beta\hbar^2}\frac{1}{2\pi\hbar}\int d\phi_nd\pi_{n}e^{i\epsilon\frac{\pi_{n}}{\hbar}(\frac{\phi_n-\phi_{n-1}}{\epsilon})}(\phi_n+\phi_{n-1})\pi_n^2.\nn
\ea
Taking account of above results and the formula
\ba
\delta_{v_n,v_{n-1}+4}-2\delta_{v_n,v_{n-1}}+\delta_{v_n,v_{n-1}-4}=\frac{1}{\pi}\int_0^\pi
db_ne^{-ib_n(v_n-v_{n-1})/2}4\sin^2(b_n),\nn
\ea
Eq.(\ref{GHamilton}) can be expressed as
\ba
&&\int
d\phi_{n}\bra{\phi_n}\bra{v_n}e^{-i\epsilon\alpha_n\hat{\Theta}}\ket{v_{n-1}}\ket{\phi_{n-1}}\nn\\
&=&\frac{1}{2\pi\hbar}\int d\phi_{n}d\pi_{n}e^{i\epsilon(\frac{\pi_{n}}{\hbar}\frac{\phi_n-\phi_{n-1}}{\epsilon})}\frac{1}{\pi}\int_0^\pi
db_ne^{-ib_n(v_n-v_{n-1})/2}\Big[1-i\alpha_n\epsilon\frac{\Delta^2}{8\kappa\gamma^2\hbar^2\phi_n}v_{n-1}\frac{v_n+v_{n-1}}{2}4\sin^2b_n\nn\\
&+&i\alpha_n\epsilon\frac{\kappa}{2\beta\hbar^2\phi_n}\left(\frac{3\hbar}{4}\right)^2
v_{n-1}\frac{v_{n}+v_{n-1}}{2}4\sin^2(b_n)+i\alpha_n\epsilon\frac{2\kappa}{i\beta\hbar^2}\left(\frac{3\hbar}{4}\right)
\pi_n v_{n-1}2i\sin(
b_n)+i\alpha_n\epsilon\frac{\kappa}{\beta\hbar^2}(\phi_n+\phi_{n-1})\pi_n^2\Big]. \nn\ea
Collecting all the above ingredients the transition amplitude can be written as
\ba
&&A_{tls}(v_f, \phi_f, \varphi_f;~v_i,\phi_i, \varphi_i)\nn\\
&=&\lim\limits_{N\rightarrow\infty}~~~~\lim\limits_{\alpha_\emph{{No}},...,\alpha_\emph{{1o}}\rightarrow\infty}
\left(\epsilon\prod\limits_{n=2}^N\frac{1}{2\alpha_\emph{{no}}}\right)\int_{-\alpha_\emph{{No}}}^{\alpha_\emph{{No}}} d\alpha_N...\int_{-\alpha_\emph{{1o}}}^{\alpha_\emph{{1o}}} d\alpha_1\times\int_{-\infty}^{\infty}d\varphi_{N-1}...d\varphi_1\left(\frac{1}{2\pi\hbar}\right)^N\int_{-\infty}^{\infty}
dp_{\varphi_N}...dp_{\varphi_1}\nonumber\\
&\times&\int_{-\infty}^{\infty}d\phi_{N-1}...d\phi_1\left(\frac{1}{2\pi\hbar}\right)^N\int_{-\infty}^{\infty}
d\pi_{N}...d\pi_{1}\sum\limits_{v_{N-1},...,v_1}~\left(\frac{1}{\pi}\right)^N\int^{\pi}_{0}db_N...db_1\nonumber\\
&\times&\prod\limits_{n=1}^{N}\exp{i\epsilon}\Big[\frac{p_{\varphi_n}}{\hbar}\frac{\varphi_n-\varphi_{n-1}}{\epsilon}
+\frac{\pi_{n}}{\hbar}\frac{\phi_n-\phi_{n-1}}{\epsilon}
-\frac{b_n}{2}\frac{v_n-v_{n-1}}{\epsilon}+\alpha_n \Big(\frac{p_{\varphi_n}^2}{\hbar^2}-\frac{\Delta^2}{8\kappa\gamma^2\hbar^2\phi_n}v_{n-1}\frac{v_n+v_{n-1}}{2}4\sin^2b_n\nn\\
&+&\frac{\kappa}{2\beta\hbar^2\phi_n}\left(\frac{3\hbar}{4}\right)^2
v_{n-1}\frac{v_{n}+v_{n-1}}{2}4\sin^2(b_n)
+\frac{2\kappa}{i\beta\hbar^2}\left(\frac{3\hbar}{4}\right)
\pi_n v_{n-1}2i\sin(
b_n)
+\frac{\kappa}{\beta\hbar^2}(\phi_n+\phi_{n-1})\pi_n^2\Big)\Big].\nn
\ea
Finally, by taking the `continuum limit' we can get a path integral formulation as
\ba
&&A_{tls}(v_f, \phi_f, \varphi_f;~v_i,\phi_i, \varphi_i)\nn\\
&=&c\int \mathcal{D}\alpha\int\mathcal{D}\varphi\int\mathcal{D}p_{\varphi}\int\mathcal{D}\phi\int\mathcal{D}\pi_{\phi}\int\mathcal{D}v\int\mathcal{D}b ~~\exp
\frac{i}{\hbar}\int_0^1d\tau \nn\\
&&\Big[p_\varphi\dot\varphi+\pi_\phi\dot{\phi}-\frac{\hbar b}{2}\dot{v}+{\hbar}{\alpha}\Big(\frac{p_\varphi^2}{\hbar^2}
-\frac{\Delta^2}{2\gamma^2\kappa\hbar^2\phi}v^2\sin^2b+\frac{2\kappa}{\beta\hbar^2\phi}\left(\frac{3\hbar}{4}\right)^2v^2\sin^2(b)
+\frac{4\kappa}{\beta\hbar^2}\left(\frac{3\hbar}{4}\right) \pi
v\sin(
b)+\frac{2\kappa}{\beta\hbar^2} \phi\pi^2 \Big)\Big],\nn
\ea
where $c$ is an overall constant.
Hence, the effective Hamiltonian constraint in the simplified model can be simply read as
\ba
C_{eff}=-\frac{\Delta^2}{2\gamma^2\kappa\hbar^2\phi}v^2\sin^2b+\frac{2\kappa}{\beta\hbar^2\phi}
\left(\frac{3\hbar}{4}\sin(b)v+\pi\phi\right)^2+\frac{p_\varphi^2}{\hbar^2}.\nn\ea
It is easy to see from above expression that the classical Hamiltonian constraint (\ref{shcm}) can be recovered from $C_{eff}$ in the large scale limit as $\sin b\rightarrow b$. Therefore the above quantum model has correct classical limit.
On the other hand, if one wants to achieve the effective Hamiltonian constraint for the original model of previous sections, the proper time of isotropic observers should be respected. Then the factor $\frac{1}{|v|}$ has to be multiplied to $C_{eff}$. We thus obtain
\ba
H_F=-\frac{\sqrt{3\Delta}}{2\gamma^2\kappa\phi}|v|\sin^2b+\frac{2\sqrt{3}\kappa}{\beta\Delta^{3/2}|v|\phi}
\left(\frac{3\hbar}{4}\sin(b)v+\pi\phi\right)^2+\frac{\Delta^{3/2}|v|}{2\sqrt{3}}\rho,\nn
\ea
where the matter density is defined by
\ba\rho=\frac{p_\varphi^2}{2\abs{p}^3}=\frac{6p_\varphi^2}{v^2\Delta^3}.\label{density}\ea
Note that the above effective Hamiltonian can also be obtained form the classical Hamiltonian (\ref{hcm}) by the heuristic replacement $b\rightarrow
\sin b$. Hence the classical Hamiltonian constraint can be recovered from the effective $H_F$ in the large scale limit.

\section{Effective equation and quantum bounce}\label{section6}

By employing the effective Hamiltonian $H_F$ and symplectic structure of Brans-Dicke cosmology, we can easily get equation of motions for
$v$ and $\phi$ respectively as
\ba
\dot{v}&=&\{v,H_F\}=\frac{2\sqrt{3\Delta}}{\hbar\gamma^2\kappa\phi}|v|\sin
(b)\cos(b)-\frac{6\sqrt{3}\kappa}{\beta\Delta^{3/2}\phi}\sgn(p)
\left(\frac{3\hbar}{4}\sin(b)v+\pi\phi\right)\cos(b),\label{vdot}\\
\dot{\phi}&=&\{\phi,H_F\}=\frac{4\sqrt{3}\kappa}{\beta\Delta^{3/2}|v|}
\left(\frac{3\hbar}{4}\sin(b)v+\pi\phi\right)\equiv\frac{4\sqrt{3}\kappa
\tilde{p}_\phi}{\beta\Delta^{3/2}|v|}. \label{phidot1} \ea
Now, let us calculate the evolution of
$\tilde{p}_\phi\equiv\frac{3\hbar}{4}\sin(b)v+\pi\phi$. It reads
\ba
\dot{\tilde{p}}_\phi&=&\{\tilde{p}_\phi,H_F\}\nn\\
&=&-\frac{3}{2}\cos(b)H_F+
H_F-\frac{\Delta^{3/2}\abs{v}}{2\sqrt{3}}\rho\nn\\
&\approx&-\frac{\Delta^{3/2}\abs{v}}{2\sqrt{3}}\rho. \label{pphi}\ea
Hence $\tilde{p}_\phi$ is a constant of motion when $\rho=0$. The combination of Eqs. (\ref{phidot1}) and (\ref{pphi}) gives
\ba
-\frac{1}{a^3}\frac{d}{dt}(\dot{\phi}a^3)&=&\frac{2\kappa}{\beta}\rho,\label{conserve}
\ea
which is as same as the classical evolution equation (\ref{seom}) for the scalar field $\phi$. However, the evolution equation corresponding to Eq.(\ref{fried}) is modified by the quantum correction, since
the combination of equations (\ref{vdot}) and
(\ref{phidot1}) gives
\ba
\left(\frac{\dot{v}}{3v}+\frac{\dot{\phi}}{2\phi}\right)^2&=&
\left[\frac{2\sqrt{\Delta}}{\sqrt{3}\hbar\gamma^2\kappa\phi}\sin
(b)\cos(b)+\frac{2\sqrt{3}\kappa}{\beta\Delta^{3/2}\phi v}
\left(\frac{3\hbar}{4}\sin(b)v+\pi\phi\right)(1-\cos(b))\right]^2\nn\\
&=&\left[\frac{2\sqrt{\Delta}}{\sqrt{3}\hbar\gamma^2\kappa\phi}\sin
(b)\cos(b)+\frac{\dot{\phi}}{2\phi
}\sgn(p)(1-\cos(b))\right]^2. \label{hubble1}\ea
On the other hand, the effective Hamiltonian constraint $H_F=0$ can be rewritten as
\ba
-\frac{3\sin^2(b)}{\kappa\gamma^2\phi\Delta}+\frac{\beta\dot{\phi}^2}{4\kappa\phi
}+\rho=0, \nn\ea
which gives
\ba \sin^2(b)=\frac{\rho_e}{\rho_c}
\label{sin1}\ea
where we defined an effective matter density $\rho_e\equiv\frac{\beta\dot{\phi}^2}{4\kappa}+\phi\rho$ and
$\rho_c\equiv\frac{3}{\gamma^2\Delta\kappa}=\frac{\sqrt{3}}{32\pi^2G^2\gamma^3\hbar}$. Note that Eq.(\ref{sin1}) guarantees the positivity of $\rho_e$. Now with
the help of Eq.(\ref{sin1}), Eq.(\ref{hubble1}) can be expressed as
\ba
\left(\frac{\dot{a}}{a}+\frac{\dot{\phi}}{2\phi}\right)^2
=\left[\frac{1}{\phi}\sqrt{\frac{\kappa}{3}\rho_e(1-\frac{\rho_e}{\rho_c})}+\frac{\dot{\phi}}{2\phi
}(1-\sqrt{1-\frac{\rho_e}{\rho_c}})\right]^2.  \label{tildeH1}\ea
Note that we also have
\ba
\rho_e
=\frac{6}{v^2\Delta^3}\left(\frac{2\kappa\tilde{p}_\phi^2}{\beta}+\phi p_\varphi^2\right):=\frac{6}{v^2\Delta^3}P_\phi^2.\nn
\ea
By using Eqs. (\ref{phidot1}), (\ref{pphi}) and (\ref{density}), we can show that $P_\phi^2$ is a constant of motion since
\ba
\dot{P}_\phi^2&=&\frac{4\kappa\tilde{p}_\phi}{\beta}\dot{\tilde{p}}_\phi+\dot{\phi} p_\varphi^2=0.\nn
\ea
Therefore, for a contracting universe, $\rho_e$ would monotonically increase while $v$ decreases. Thus, when $\rho_e=\rho_c$, we have
$\cos(b)=\sqrt{1-\frac{\rho_e}{\rho_c}}=0$. Then, from Eq.(\ref{vdot}), we can easily get $\dot{v}=0$, which implies a quantum bounce
happened at that point. To see this is really the case, we can calculate $\ddot{v}$ by taking the Poisson bracket of Eq.(\ref{vdot}) with the effective Hamiltonian $H_F$ as
\ba
\ddot{v}=-\frac{4\sqrt{3\Delta}}{\hbar^2\gamma^2\kappa\phi}|v|\left(\cos^2b-\sin^2
b\right)-\frac{12\sqrt{3}\kappa}{\hbar\beta\Delta^{3/2}\phi}\sgn(p)
\left(\frac{3\hbar}{4}\sin(b)v+\pi\phi\right)\sin(b)+\frac{9\sqrt{3}\kappa}{\beta\Delta^{3/2}\phi}\abs{v}\cos^2(b).\label{ddotv}
\ea
Now we consider the evolution at the point $\rho_e=\rho_c$. Since we have $\cos(b)=0$ and $\sin(b)=\sgn(p)$, taking account of Eq.(\ref{phidot1}), Eq.(\ref{ddotv}) becomes
\ba
\ddot{v}\mid_{\rho_e=\rho_c}=\frac{4\sqrt{3\Delta}}{\hbar^2\gamma^2\kappa\phi}|v|-\frac{3\dot{\phi}}{\hbar\phi}\abs{v}=\frac{6\abs{v}}{\hbar\phi}
\left(\sqrt{\frac{\kappa\rho_c}{3}}-\frac{\dot{\phi}}{2}\right).\label{ddotv2}
\ea
To simplify the discussion, we now consider the vacuum situation. Then we have $\rho_e=\frac{\beta\dot{\phi}^2}{4\kappa}$, and hence Eq.(\ref{ddotv2}) becomes
\ba
\ddot{v}\mid_{\rho_e=\rho_c}=\frac{6\abs{v}}{\hbar\phi}\sqrt{\frac{\kappa\rho_c}{3}}\left(1-
\sqrt{\frac{3}{\beta}}\right)\neq 0,\label{bounce}
\ea
where we used the fact that the Brans-Dicke coupling parameter $\omega=0$ was ruled out by the solar system experiments \cite{will,will1}.
To justify the above auguments, let us consider the vacuum solution of this cosmological model. In this case, Eq.(\ref{conserve}) becomes
\begin{equation}
\ddot{\phi} + 3\frac{\dot{a}\dot{\phi}}{a} = 0.\label{vacuum}
\end{equation}
Plugging Eq.(\ref{vacuum}) and $\rho_e = \frac{\beta\dot{\phi}^2}{4\kappa}$ into Eq.(\ref{tildeH1}), we get
\begin{equation}
\left(\frac{\dot{\phi }}{2\phi } - \frac{\ddot{\phi }}{3\dot{\phi }}\right)^2 =  \left(\frac{\dot{\phi }}{2\phi }\right)^2\left(\left(\sqrt{\frac{\beta }{3}}-1\right)\sqrt{1 - \frac{\beta }{4\rho _{c}}\dot{\phi }^2} + 1\right)^2 \ .\label{solution}
\end{equation}
Let us consider $\dot{\phi}$ as a function of $\phi$, i.e., $\dot{\phi}\equiv f(\phi)$. This implies $\ddot{\phi}=f'f$, where $f'\equiv\frac{df}{d\phi}$. Assuming $\frac{\dot{a}}{a} + \frac{\dot{\phi }}{2\phi }\geq 0$ in Eq.(\ref{tildeH1}), from Eq.(\ref{solution}) one finds
\begin{equation}
\frac{2\phi }{3}f' = - f\left(\sqrt{\frac{\beta }{3}}-1\right)\sqrt{1 - \frac{\beta }{4\rho _{\text{c}}}f^2}\ .
\end{equation}
The solution of this equation goes as follows,
\begin{equation}
f = \dot{\phi} = \frac{4\sqrt{\frac{\rho _{\text{c}}}{\beta }}\left(\frac{\phi}{\phi_{cr}}\right) ^{\frac{3+\sqrt{3\beta } }{2}} }{\left(\frac{\phi}{\phi_{cr}}\right) ^3+\left(\frac{\phi}{\phi_{cr}}\right) ^{\sqrt{3\beta }}} \ ,
\end{equation}
where $\phi_{cr}$ is the value of the field at the moment of the bounce. Therefore we obtain
\begin{equation}
\frac{\dot{a}}{a} = -\frac{1}{3}f' = \frac{2 \left( \sqrt{\frac{\beta }{3}}-1\right)\sqrt{\frac{\rho _{\text{c}}}{\beta }} \left(\frac{\phi}{\phi_{cr}}\right) ^{\frac{1}{2}\left(1+ \sqrt{3\beta }\right)} \left(\left(\frac{\phi}{\phi_{cr}}\right) ^{\sqrt{3\beta }}-\left(\frac{\phi}{\phi_{cr}}\right) ^3\right) }{\phi_{cr}\left(\left(\frac{\phi}{\phi_{cr}}\right) ^3+\left(\frac{\phi}{\phi_{cr}}\right) ^{\sqrt{3\beta }}\right)^2}\ , \label{eq:HubbleAnBD}
\end{equation}
The special case of this solution is $\beta = 3$, which corresponds to $\omega = 0$. In such a case one gets a constant Hubble parameter $\mathcal{H}\equiv\frac{\dot{a}}{a}=0$, which is unphysical and coincides with the result of Eq.(\ref{bounce}).

\begin{figure}[h]
\centering
\includegraphics*[height=5cm]{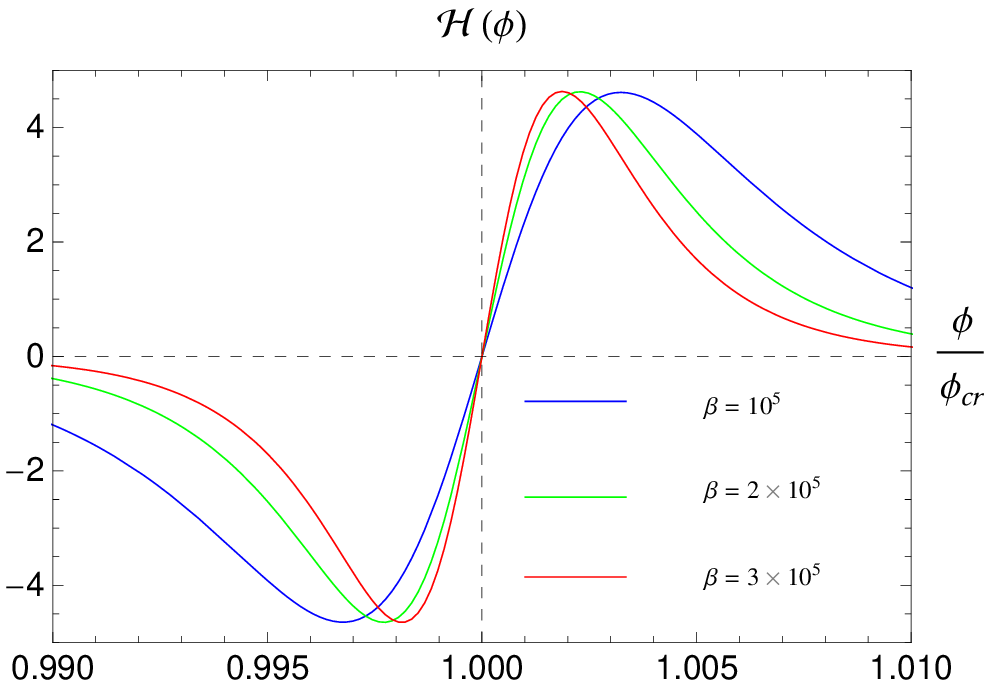}
\hspace{0.4cm}
\includegraphics*[height=5cm]{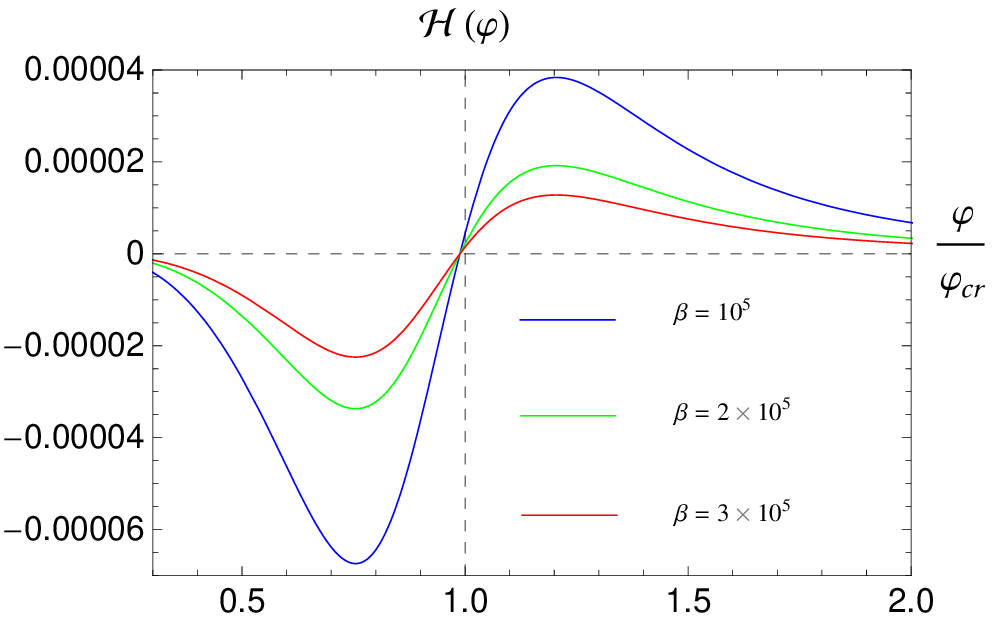}
\caption{\it Left and right panels present the evolution of the Hubble parameter in the Planck units as a function of the Brans-Dicke field (left panel, vacuum solution) or the massless scalar field (right panel, massless scalar field domination) for realistic values of $\beta$. Initial conditions are chosen to be: $\phi_{cr}=1$ (left panel) and $\frac{\dot{\phi}_{cr}}{\dot{\varphi}_{cr}}=-\frac{\varphi_{cr}}{\beta}$, $\phi_{cr}-\frac{\varphi_{cr}^2}{\beta}-\frac{\dot{\phi}_{cr}}{\dot{\varphi}_{cr}}\varphi_{cr}=M_{pl}$ (right panel)}
\label{fig:HBD}
\end{figure}

Except for the unphysical case, the evolution of the Hubble parameter for $\omega\sim 10^5$ is presented at the left panel of the Fig.\ref{fig:HBD}. It should be noted that, the assumption which we made in order to obtain these solutions is satisfied for any values of $\beta$, $\rho_{c}$ and $\phi_{cr}$, as long as $\phi>0$ (for $\beta\in (0,12)$) or $\phi > \phi_{cr} \left(1 - \frac{2\sqrt{3}}{\sqrt{\beta }}\right)^{1/(\sqrt{3\beta }-3)}$ (for $\beta>12$). Since $\phi = \phi_{cr}$ is always within this range, the solution (\ref{eq:HubbleAnBD}) covers the bounce for any values of parameters of the model.
The similar analysis could also be performed for the effective Brans-Dicke cosmology with a massless scalar field $\varphi$. We present the existence of the bounce in this case at the right panel of Fig.\ref{fig:HBD} by the evolution of the Hubble parameter as a function of $\varphi$, which in this case may play a role of a time variable.
Hence, just as the LQC case of GR, the big bang singularity of classical Brans-Dicke cosmology can also be avoided by its loop quantization.

We end up this section with following
two remarks on the effective equation (\ref{tildeH1}).
(i) In the special case of $\phi=1$, we have $\rho_e=\rho$ and $\dot{\phi}=0$. Then Eq.(\ref{tildeH1}) reduces to the well-known
effective Friedman equation of LQC as
\ba
\left(\frac{\dot{a}}{a}\right)^2
=\frac{\kappa}{3}\rho(1-\frac{\rho}{\rho_c}). \nn\ea
(ii) In the classical limit when $\rho_e\ll\rho_c$, the
$\frac{\rho_e}{\rho_c}$ terms in Eq. (\ref{tildeH1}) can be neglected, and hence it reduces to the
 evolution equation (\ref{fried}) of classical Brans-Dicke cosmology as
\ba
\left(\frac{\dot{a}}{a}+\frac{\dot{\phi}}{2\phi}\right)^2
=\frac{1}{\phi^2}\frac{\kappa}{3}\rho_e
=\frac{\beta\dot{\phi}^2}{12\phi^2}+\frac{\kappa\rho}{3\phi}.\nn\ea

\section{concluding remarks}\label{section7}

To summarize the results in previous sections, we first studied the spatially flat FRW model of Brans-Dicke theory. It turns out that,
although the scalar field is non-minimally coupled, it can still be
treated as an emergent time variable. Hence, in Brans-Dicke cosmology an internal time may come from the gravity rather than an extra matter field. This model is then successfully quantized by the nonperturbative loop quantization approach with the Brans-Dicke coupling parameter $\omega\neq-\frac32$.  The Hamiltonian constraint is successfully quantized in this model. Due to the polymer-like quantization and the
non-vanishing minimal area, the classical differential equation
which represents cosmological evolution is now replaced by quantum
difference equation. In addition, we use the timeless path integral
formalism and the simplified treatment to derive an effective Hamiltonian of loop quantum
Brans-Dicke cosmology. The same expression could also be obtained if we took the
heuristic replacement of $\ct\rightarrow
\frac{\sin(\mubar\ct)}{\mubar}$ in the classical Hamiltonian constraint. Hence the quantum theory has correct classical limit. Furthermore, we use this effective
Hamiltonian to get the effective dynamical equations of the theory, which lay a foundation for the phenomenological investigation to possible quantum gravity effects in cosmology. Our analysis indicates that the classical big bang
singularity is again replaced by a quantum bounce in loop quantum Brans-Dicke
cosmology. This result strengthens our confidence that the existence of quantum bounce is a universal
feature of loop quantum cosmological models.

There is also interesting situation in loop quantum
Brans-Dicke cosmology, which does not exist in the LQC of GR. Since the scalar field $\phi$ of Brans-Dicke gravity can play the role of emergent time, there exists a meaningful vacuum evolution in loop quantum Brans-Dicke
cosmology. As shown in section V, in this case the quantum bounce still exists even without
extra matter field.
It should be noted that there are many aspects of the loop
quantum Brans-Dicke cosmology which deserve further investigating.
For examples, it is still desirable to confirm the effective equations of
loop quantum Brans-Dicke cosmology from canonical perspective. To confirm the universality of the quantum bounce, we need to generalize our scheme
to other modified gravity theories, such as $f(R)$ theories and general scalar-tensor
theories. Moreover, since our effective equations laid a foundation for the phenomenological investigation to possible quantum gravity effects in cosmology, we also would like to further study the cosmological
perturbation theory and inflation scenario under our framework of
loop quantum Brans-Dicke cosmology. We leave all these interesting
topics for future study.

\begin{acknowledgements}
This work is supported by NSFC (No.10975017, No.11235003 and No.11275073)  and
the Fundamental Research Funds for the Central University of China
under Grant No.2012ZZ0079. M.A. would also like to acknowledge China Postdoctoral
Science Foundation for financial support.

\end{acknowledgements}



\begin{thebibliography}{99}



\bibitem{Ro04} C. Rovelli, {\it Quantum Gravity,} (Cambridge University Press, 2004).

\bibitem{Th07} T. Thiemann, {\it Modern Canonical Quantum General Relativity,} (Cambridge University
Press, 2007).


\bibitem{As04}A. Ashtekar and J. Lewandowski, {\it Background independent quantum gravity: A
status report,} Class.Quant.Grav. {\bf21}, R53 (2004).

\bibitem{Ma07} M. Han, W. Huang, and Y. Ma, {\it Fundamental structure of loop quantum gravity,}
 Int. J. Mod. Phys. D {\bf16}, 1397 ,(2007).


\bibitem{Zh11} X. Zhang and Y. Ma, {\it Extension of loop quantum gravity to $f(R)$
theories,} Phys. Rev. Lett. {\bf 106}, 171301 (2011).


\bibitem{Zh11b} X. Zhang and Y. Ma, {\it Loop quantum f(R) theories,} Phys. Rev. D {\bf 84}, 064040 (2011).


\bibitem{ZM12a} X. Zhang and Y. Ma, {\it Loop quantum Brans-Dicke
theory,} J. Phys.: Conf. Ser. {\bf 360}, 012055 (2012).


\bibitem{ZM11c} X. Zhang and Y. Ma,  {\it Nonperturbative loop
quantization of scalar-tensor theories of gravity,} Phys. Rev. D {\bf 84}, 104045 (2011).

\bibitem{Ma12a} Y. Ma, {\it Extension of loop quantum gravity to metric theories beyond general relativity,} J. Phys.: Conf. Ser. {\bf 360}, 012006 (2012).

\bibitem{LQC5}
A. Ashtekar, M. Bojowald, and J. Lewandowski, {\it Mathematical structure of loop quantum cosmology,} Adv. Theor. Math.
Phys. \textbf{7}, 233 (2003).

\bibitem{Boj}
M. Bojowald, {\it Loop quantum cosmology,} Living Rev. Relativity \textbf{8}, 11 (2005).

\bibitem{AS11} A. Ashtekar, P. Singh, {\it Loop quantum cosmology: A status
report,}  Class. Quant. Grav. {\bf28}, 213001 (2011).

\bibitem{APS3} A. Ashtekar, T. Pawlowski, P. Singh,  {\it Quantum nature of the big bang: Improved
dynamics,}  Phys. Rev. D {\bf 74}, 084003 (2006).

\bibitem{BD} C. Brans and R. H. Dicke, {\it Mach's principle and a relativistic theory of
gravitation,} Phys. Rev. {\bf 124}, 925 (1961).

\bibitem{Greenstein1}G. S. Greenstein, {\it Brans-Dicke cosmology, I,} Astrophys. Letter. {\bf 1},
139 (1968).

\bibitem{Greenstein2}G. S. Greenstein, {\it Brans-Dicke cosmology, II,} Astrophysics and Space Science {\bf 2}, 155 (1968).

\bibitem{BD2}J. D. Anderson and J. R. Morris, {\it Brans-Dicke theory and the Pioneer
anomaly,} Phys. Rev. D  {\bf 86}, 064023 (2012).

\bibitem{Banerjee} N. Benerjee and D. Pavon, {\it Cosmic acceleration without quintessence,} Phys. Rev. D {\bf63}, 043504
(2001).

\bibitem{Sen}S. Sen and A. A. Sen, {\it Late time acceleration in Brans-Dicke cosmology,} Phys. Rev. D \textbf{63}, 124006(2001).

\bibitem{Qiang}L. Qiang, Y. Ma, M. Han and D. Yu, {\it 5-dimensional Brans-Dicke theory and cosmic acceleration,} Phys. Rev. D {\bf71},  061501(R)
(2005).

\bibitem{DB}S. Das, N. Banerjee, {\it Brans-Dicke scalar field as a
chameleon,} Phys. Rev. D {\bf 78}, 043512 (2008).

\bibitem{FT}A. D. Felice, S. Tsujikawa, {\it Generalized Brans-Dicke
theories,} JCAP {\bf 1007}, 024 (2010).

\bibitem{BD1} Y. Bisabr, {\it Cosmic acceleration in Brans-Dicke
cosmology,} Gen. Rel. Grav. {\bf44}, 427 (2012).

\bibitem{01}J. Friemann, M. Turner, D. Huterer, {\it Dark energy and the accelerating
Universe,}
 Ann. Rev. Astron.
Astrophys. {\bf46}, 385 (2008).

\bibitem{will} C. M. Will,  {\it The confrontation between general relativity and
experiment,} Living Rev. Relativity {\bf 9}, 3 (2006).

\bibitem{will1}C. M. Will, {\it Theory and Experiment in Gravitational
Physics,} (Cambridge University Press, 1993).

\bibitem{Ash-view}A.~Ashtekar, {\it Loop quantum cosmology: An overvie,} Gen. Rel. Grav. {\bf41}, 707 (2009).

\bibitem{ACS}A. Ashtekar, A. Corichi, and P. Singh, {\it Robustness of key features of loop quantum
cosmology,} Phys. Rev. D  {\bf 77}, 024046 (2008).

\bibitem{Taveras}
V. Taveras, {\it Corrections to the Friedmann equations from loop quantum gravity for a universe with a free scalar field,} Phys. Rev. D \textbf{78}, 064072 (2008).

\bibitem{DMY}
Y. Ding, Y. Ma and J. Yang, {\it Effective scenario of loop quantum cosmology,} Phys. Rev. Lett. \textbf{102}, 051301
(2009).

\bibitem{YDM}
J. Yang, Y. Ding and Y. Ma, {\it Alternative quantization of the Hamiltonian in loop quantum cosmology,} Phys. Lett. B \textbf{682}, 1 (2009).

\bibitem{Boj11}
M. Bojowald, D. Brizuela, H. H. Hernandez, M. J. Koop, H. A.
Morales-Tecotl, {\it High-order quantum back-reaction and quantum cosmology with a positive cosmological constant,} Phys. Rev. D \textbf{84}, 043514 (2011).

\bibitem{ACH102}
A. Ashtekar, M. Campiglia, A. Henderson, {\it Loop quantum cosmology and spin foams,} Phys. Lett. B \textbf{681}, 347 (2009); {\it

Casting loop quantum cosmology in the spin foam paradigm,} Class. Quant. Grav. \textbf{27}, 135020 (2010); {\it Path integrals and the WKB approximation in loop quantum cosmolog,} Phys. Rev. D \textbf{82}, 124043 (2010).

\bibitem{QHM}
L. Qin, H. Huang and Y. Ma, \emph{Path integral and effective
Hamiltonian in loop quantum cosmology}, Gen. Rel.
Grav. in press.

\bibitem{QDM}
L. Qin, G. Deng and Y. Ma, {\it Path integral and effective Hamiltonian in loop quantum cosmology,} Commun. Theor. Phys. \textbf{57}, 326
(2012).

\bibitem{QM1}
L. Qin and Y. Ma, {\it Coherent state functional integrals in quantum cosmology,} Phys. Rev. D \textbf{85}, 063515 (2012).

\bibitem{QM2}
L. Qin and Y. Ma, {\it Coherent state functional integral in loop quantum cosmology: Alternative dynamics,} Mod. Phys. Lett. \textbf{27}, 1250078 (2012).


\end{thebibliography}
\end{document}